\documentclass[pre,nofootinbib,twocolumn,floatfix]{revtex4}
\usepackage{graphicx}
\bibpunct[, ]{[}{]}{,}{a}{,}{,}
\ifnum\lefthyphenmin<2\lefthyphenmin=2\fi
\ifnum\righthyphenmin<2\righthyphenmin=2\fi
\newcommand{\be}{\begin{equation}}
\newcommand{\ee}{\end{equation}}
\newcommand{\bea}{\begin{eqnarray}}
\newcommand{\eea}{\end{eqnarray}}

\begin{document}

\title{Content based network model with duplication and divergence}

\author{Yasemin \c Seng\"un and Ay\c se Erzan}
\affiliation{Department of Physics, Faculty of Sciences and 
Letters\\
Istanbul Technical University, Maslak 34469, Istanbul, Turkey}


\date{\today}
\begin{abstract}
 
We construct a minimal content-based realization of the  duplication and divergence model of genomic networks introduced by
 Wagner [A. Wagner, Proc. Natl. Acad. Sci. {\bf 91}, 4387 (1994)] and investigate the
scaling properties of the directed degree distribution and clustering coefficient.  We find that the content based
network exhibits crossover between two scaling regimes, with log-periodic oscillations for large degrees. These
features are not present in the original gene duplication model, but inherent in the content based model of Balcan and
Erzan. The scaling exponents $\gamma_1$ and $\gamma_2=\gamma_1-1/2$ of the Balcan-Erzan model turn out to be robust 
under duplication and point mutations,
but get modified in the presence of  splitting and merging of strings.  The clustering coefficient as a function of the
degree, $C(d)$, is found, for the Balcan-Erzan model, to behave in a way qualitatively similar to the out-degree
distribution, however with a very small exponent $\alpha_1= 1-\gamma_1$ and an envelope for the oscillatory part, which is essentially flat, thus $\alpha_2= 0$.  Under duplication and mutations 
including splitting and merging of strings, $C(d)$ is found to decay exponentially.


\end{abstract}

\maketitle

\section{Introduction}

Biological networks have recently been the subject of many theoretical studies~\cite{Barabasi1999,Maslov,Barabasi,Dorogovtsev}. The gene-duplication model  introduced by Wagner~\cite{Wagner1,Wagner,Satorras,genenetworks} for protein interaction provides a biologically motivated stochastic mechanism for the emergence of a small-world, scale free network~\cite{Watts-Strogatz}, with the topological properties  depending on just one parameter.  This parameter may be adjusted to obtain values of the scaling exponent for the degree distribution, which may be compared  with those reported in the literature for the genomic networks. There is a wealth of evidence pointing to the fact that a major portion of the genomic network in many organisms in fact originate from the duplication of transcription factors (TF) and target genes~\cite{Alberts,Lewin,Teichmann1,Teichmann2}.

Wagner's mathematical model~\cite{Wagner1,Satorras} involves nodes, corresponding to genes, and edges, corresponding to genomic interactions, connecting the nodes. Duplication of the nodes results in the inheritance of all of their previous connections.  It is postulated that there exists a finite mutation rate which causes the edges to be removed,  with some fixed probability.  In this minimal model, once a number of interactions are in place, evolution by gene duplication and mutations generically drives the network to a scale free topology.  Natural selection can then act on this spontaneously arising scale free network. 

The purpose of the present study is to  associate the point-like nodes of the Wagner model with binary strings of different lengths, and investigate the behavior of the network under fixed rates of duplication and mutation of these strings. For this, one must {\em i)} postulate a rule for the connectivity of the network, {\em ii)} specify the particular ways in which duplications and mutations are to be affected. 

The  connectivity rule which we will adopt was originally proposed by Balcan and Erzan~\cite{Balcan-Erzan} for a content-based null model for genomic networks. In this model (henceforth, the BE model) the connectivity  is decided by the  similarity of the strings associated with each node.   This model is striking in that the content based connectivity rule  gives rise to a spontaneous self-organization of the genome into a network with a definite scaling behavior~\cite{Balcan-Erzan,mungan}. The suggestion is that the assembly of a linear code of sufficient length already provides the possibility for  complex inter-genomic interactions based on homologies, which may later be fine tuned by natural selection. 

In the present paper we will combine the inputs from the two models, the Wagner model and the BE model, to investigate how the content based interaction model evolves under duplications and mutations. We will concentrate on the out-degree distribution to characterize the topology of the network. We will also give an {\em ab initio}  derivation of the clustering coefficient for the BE model, and investigate how its scaling behavior~\cite{Milo,Vazquez} changes under duplications and mutations.

It should be mentioned that there have recently been several studies of models of gene regulatory networks on ``Artificial Genomes''(AG)~\cite{Reil} based on various alphabets and matching rules. The Wagner model has been simulated with strings mimicking genes, promoter sequences and transcription factors,  which evolve under duplication and divergence due to mutations ~\cite{Wiles,vannoort,Banzhaf}. These derivative models employ different degrees of ``realism'' in their schematization of the interactions involved. In this paper we will stick to the simplest possible rules of perfect matching between binary strings associated with each node, for the establishment of the interaction matrix.  Moreover we will not make any distinctions between regulatory sequences, genes, or TFs, keeping the model to a minimum, so that it may eventually be amenable to analytical treatment. 

In section 2, we outline the original BE model for completeness, and give a summary of its properties. In section 3 we define an out-clustering coefficient, and report on our analytical results for its scaling behavior as a function of the  out-degree. In section 4, we introduce the  protocols for duplication and mutation, which we have adopted in the present study.  In section 5, we report the results of our simulations.  Section 6 provides a discussion with an overview of other content based models of duplication and divergence.

\section{The BE model} 

The BE network~\cite{Balcan-Erzan} is defined via the following rules.
\begin{enumerate}
\item
Consider an alphabet consisting only of the symbols $\{0,1,2,\ldots r\}$, with the character ``r'' always signifying the delimiter between successive words. For convenience, we assume that an ``$r$'' has also been placed 
at the $0^{th}$ and $(L+1)^{th}$ positions of the whole sequence. (In the original BE paper, and in most of the rest of this paper, $r=2$.)
\item Form a linear string  $\cal{C}$ (our ``Artificial Genome'') of length $L$, by randomly choosing letters  from  this alphabet, according to a predetermined distribution. We will think of this string as a one dimensional lattice and speak about the ``$k$th site'' or the ``$k$th member of the sequence,'' interchangeably.
\item Associate the substrings $G_i$ intervening between the $i$th and $i+1$st delimiters, with the $i+1$st node of a graph. (Thus the strings $G_i$ are formed from the characters $\{0,1,\ldots r-1\}$, therefore an alphabet of length $r$).
\item The connectivity rule: Postulate a directed edge to exist between the  nodes $(i,j)$  
if and only if the string $G_i$
associated with the $i$th node occurred at least once in the random 
word $G_j$ associated with the $j$th node, i.e., $G_i \subseteq G_j$ .  This ``inclusion relation'' determines the connectivity matrix $w_{ij}$. Thus
\begin{equation}
 w_{ij} = 1 \;\;\;\;{\rm if} \;\;\;\;  G_i \subseteq G_j 
\label{connmatrix}
\end{equation}
and is zero otherwise. Note  $w_{ij} \neq w_{ji}$ unless the lengths of the strings associated with the nodes $i,j$ happen to be equal. Clearly $w_{ij}=0$ for $l_j < l_i$. 
\end{enumerate}

In the two previous papers~\cite{Balcan-Erzan,mungan}, we chose the following distribution function for the independently and identically distributed random variables $x$ occupying each of the positions along the string $\cal{C}$,
\begin{equation}
P(x) = p \delta(x-r) + (1-p)\frac{1}{r}\sum_{m=0}^{r-1}\delta(x-m)\;\;.
\label{probdist1}
\end{equation}
 Note that the resulting set of words obey the following sum rules,
\begin{equation}
\label{sum_rules}
\sum_i{\ell_i} = L-N\ , \ \ \ \sum_{\ell}{n_\ell} = N\ ,
\end{equation}
and moreover,
\begin{equation}
\label{averages}
\langle \ell \rangle = p^{-1} -1, 
 \ \ \ \langle n_\ell \rangle = Lp^2q^\ell, \  \ \ 
\langle N \rangle = Lp .
\end{equation}
where $n_l$ is the number of nodes with associated words of length $l$. Note that the notation $\langle\ldots\rangle$ signifies averages over many different realizations of $\cal{C}$.

The topological properties of this model have been extensively reported in
two previous papers~\cite{Balcan-Erzan,mungan}, so we will only give a
brief summary here.  For the network to consist of a large connected
cluster, the frequence of the delimiters should be large enough; equivalently, the average word length should not be too large. This introduces a threshold for $p$, such that the network is connected for $p\ge p_c(L)$, where $p_c$
decreases with $L$~\cite{Balcan-Erzan}. Here we will assume $ p_c < p \ll 1 $.
We checked that the properties of the network depended only very weakly on
$p$ as long as it is small.  The connected cluster was
found~\cite{Balcan-Erzan} to be of the extremely small-world
type~\cite{Barabasi1999,Barabasi,Dorogovtsev}, with the supremum of the
smallest distance between any given pair of nodes being independent of the
size of the network, and an average clustering coefficient which is much
larger than that expected for Erd\"os-Renyi random networks~\cite{Erdos}.

We have performed both simulations~\cite{Balcan-Erzan} and analytical calculations~\cite{mungan} to determine the {\em expected } out-degree distribution, namely the average degree distribution computed over a large ensemble of independent realizations of the sequences $\cal{C}$ of identical length $L$.
We find that those nodes with high out-degrees $d$ are associated with short strings, of length $l = 1,2,3,\ldots$, and that for small $p$ and finite $L$  the degree distribution is Poisson about a mean 
\begin{equation}
d_l = {N \over p+q z^l} (qz)^l
\label{dl}
\end{equation}
for each $l$.  We have defined $z=1/r $ and $q=1-p$. The width of these peaks  are given by 
\begin{equation}
\sigma_l^2 =  d_l\;\;,
\end{equation}
for $p\ll 1$ and for finite $L$.
The spacing between the peaks shrinks exponentially with $l$, while their width shrinks exponentially with $l/2$, so that for $l$ larger than  a critical value, the peaks merge to yield a smooth distribution. Thus there are two distinct regimes in the out-degree distribution.  For relatively small $l$, where there is a log-periodic oscillatory behavior with discrete peaks, the envelope of the log-periodic oscillations $E(d)$ behaves as 
\begin{equation}
E(d) = \sqrt{{N \over 2 \pi}}  (d/N)^{-\gamma_2} \label{envelope}
\end{equation}
with 
\begin{equation}
\gamma_2 = {1\over 2} {\ln z - \ln q \over \ln z + \ln q}  \simeq  
{1\over 2} \left( 1+{p \over \ln z}\right) \;\;\;.
\label{gamma2}
\end{equation}
It should be noted that if we allow $L \to \infty$ (and thereby $N\to \infty$), the finite size scaling corrections no longer apply.  Then, the variance of the peaks is given by 
\begin{equation}
\sigma_l^2 =  d_l {p(1-z^l) \over 1-q(1-z^l)^2}\;\;\;,
\end{equation}
for small $l$, which is smaller than the finite size result, and the envelope no longer decays with very large $d$, but instead grows, becoming higher and sharper for smaller $l$.~\cite{mungan}  

As $N$ or $L$ are allowed to increase without bound, we find that 
\begin{equation}
E^\infty (d) = E(d) (1+a_1 d^{\delta_1} + a_2 d^{\delta_2}\ldots)
\end{equation}
where the exponents are 
\begin{equation}
\delta_1 = {\ln z \over \ln z - \ln q}\;\;\; \delta_2 = 2 \delta_1\;\;.
\end{equation}
Note that the area under the  $l=1$ peak in fact counts the total number of non-null nodes in the network.

In the large $l$, small $d$  regime, the distribution scales with a different exponent,  
\be
P(d) \sim d^{-\gamma_1} \;\;,
\label{P(d)}
\ee
for which our analytical calculations yield
\begin{equation}
\gamma_1 = {1\over 2} + \gamma_2\;\;\;.
\label{gamma1}
\end{equation}

The in-degree distribution is more localized, and  also displays a discrete structure for small degrees, while the large degree tail  decays like a Gaussian.  

\section{The clustering coefficient}

The  clustering coefficient $C(d)$~\cite{Milo,Vazquez} measures the probability that two nodes which share a neighbor of 
degree $d$ are connected between themselves, and is  defined as
\begin{equation}
C(d) = \langle {1 \over n(d)} \sum_i^{n(d)} { 2\epsilon_i \over d(d-1)}\rangle  \;\;,
\label{clustering}
\end{equation}
where $n(d)$ is the number of nodes with degree $d$, the sum is over all nodes of degree $d$, and $\epsilon_i$ is the
number of edges which connect the neighbors of the $i$th node of degree $d$. The brackets again signify averaging over an ensemble of independent realisations. Clearly, if we multiply this quantity
with $ \left( \begin{array}{ll} d \\ 2 \end{array} \right)$ we obtain the average number of triangles which which pass
through nodes of degree $d$.  Therefore $C(d)$  is of interest in discussing the prevalence of different
motifs~\cite{Milo} within a given network.  It has been conjectured that this quantity scales with $d$, such that $C(d) \sim d^{-\alpha},$ with $\alpha$ being called the ``hierarchical exponent''~\cite{Milo,Vazquez}.
 
We find that a more interesting quantity in our context is the clustering coefficient defined with respect to the out-neighbors of a node with out-degree $d$. We have,
\begin{equation}
C_{\rm out}(d) = \langle {1 \over n_{\rm out}(d)} \sum_i^{n_{\rm out}(d)} { 2\epsilon_i \over d(d-1)}\rangle \;\;.
\label{out-cluster}
\end{equation}
Note that, due to the transitivity of the connectivity rule (\ref{connmatrix}), $w_{ab} w_{bc} = 1$ implies 
$w_{ac}=1$. This provides an important simplification in the computation of $C_{\rm out}(d)$.  From now on we will
drop the index ``out,'' and mean (\ref{out-cluster}) by $C(d)$.

We have,
\begin{equation}
C(d) = \sum_{k_2}^L \sum_{k_1}^{k_2} \sum_{l=1}^{k_1} P(l\vert d)\, p(k_1,k_2,\vert l)\, \tilde  p(k_1) \, \tilde 
p(k_2)\;\;.
\label{out-cluster-1}
\end{equation}
Here, $P(l\vert d)$ is the probability that a node with degree $d$ has the length $l$, and $\tilde p(k_i)$ is the
probability that a neighbor of such a node has the length $k_i$. Using $P(d\vert l)$, the probability that a string of
length $l$ has the out-degree $d$, we may write,
\begin{equation}
P(l \vert d) = {P(d \vert l) p(l) \over P(d)}\;\;,
\label{out-cluster-2}
\end{equation}
where $p(l)= n_l/N $ is the probability of encountering a string of length $l$, $P(d)$ is the probability of a node of out-degree $d$, and
\begin{equation}
\tilde p(k_i)= {p(k_i) \over \sum_{k = l}^L p(k)} \;\;.
\end{equation}
Finally, $p(k_1,k_2,\vert l)$ is the probability that a string of length $k_1$ is found inside a string of length $k_2 \ge k_1$, given that the same string of length $l\le k_1$ is found in both.  Note that, in obvious notation,
\begin{equation}
p(k_1,k_2,\vert l) \, p(l \in k_1, l\in k_2) = p(k_1 \in k_2, l\in k_1, l\in k_2) \;\;.
\end{equation}
Due to the transitivity relation stated above, the right hand side  reduces to $p(k_1 \in k_2, l\in k_1)$.  Using the fact that the joint probability on the left hand side must factorize,  since each string is constituted independently, we have,

\begin{equation}
p(k_1,k_2,\vert l) = p(k_1 \in k_2) / p(l\in k_2)\;\;.
\label{ratio}
\end{equation}
The quantities $p(l \in k) \equiv p(l,k)$ are simply the probability that a random string of length $l$ is found in a string of length $k \ge l$.  These quantities  have been computed in ~\cite{mungan} as
\begin{equation}
p(l,k) = 1 - \left ( 1 - r^{-l} \right )^{k-l+1}
\end{equation}
for strings drawn randomly from an alphabet of length $r$.  We also have from ~\cite{mungan} that, for small $d$ (large $l$), 
\begin{equation}
P(d\vert l) = { d_l^d e^{-d_l} \over d!}
\end{equation}
where $d_l$ is given by (\ref{dl}) and $P(d)$ by (\ref{P(d)}).  

We would now like to find the scaling behavior of $C(d)$. For small $d$, substituting all the relevant quantities in
(\ref{out-cluster-1}), one finds that the $l$-sum may be performed using a saddle-point approximation, and is
independent of the upper limit, so that the subsequent sums over $k_1, \; k_2$ lead to constants. After some algebra we find,
\begin{equation}
C(d) \simeq {\rm const.}\; 
d^{\,-(1-\gamma_1)} \sim d^{\,-\alpha_1}\;\;,
\label{alpha1}
\end{equation}
where
\begin{equation}
\alpha_1  = 1-\gamma_1 \simeq {p \over p + \ln r}\;\;,
\end{equation}
with the approximation being valid for small $p$, giving the numerical value, for $r=2$,  $p=0.05$, of $\alpha_1 = 0.07$.

In the region of large $d$ (small $l$), we have already seen that the distribution of nodes with out-degree $d$ breaks up into discrete peaks.  For each $l = d_l^{-1}(d)$, where the inverse function is meant by the power $-1$, we have the height of the peak in the attendant $C(d)$ distribution to be 
\begin{equation}
C[l(d)] = \sum_{k_2=l}^L \sum_{k_1=l}^{k_2}   p(k_1,k_2,\vert l) \, \tilde  p(k_1) \, \tilde p(k_2)\;\;.
\label{out-cluster-3}
\end{equation}
It is actually possible to sum this series to lowest order in $(k_1 - l )/l$ and the leading term is simply independent of $l$, and therefore of $d$. Thus, the envelope of these peaks behaves as  $ \sim d^{\,-\alpha_2} \sim {\rm const.}$, or, in other words, 
\begin{equation}
\alpha_2 \,= 0\;\;.
\end{equation}
These results reveal that the out-clustering coefficient is essentially independent of $d$ for all practical purposes, for both small and large out-degrees $d$.

The simulations results completely agree with these findings, as will be shown in Section 5.

\section{Content based model for the evolution of genomic networks}

In this section we define our network model with duplication and
divergence, based on the original BE model (Section 2), with the different
algorithms for evolution under duplication and various kinds of mutations.  
To avoid confusion, apart from standard terminology, we will reserve the
term ``sequence'' for the complete sequence $\cal{C}$ of length $L$.  The
subsequences $G_i$ of $\cal{C}$, which appear between the delimiter marks, we
will call ``strings,'' or ``words.'' In the rest of the paper, we will
restrict the model to $r=2$, so that the strings (words) to be associated
with the nodes of the network are binary strings whose elements are $\in
\{0,1 \}$, and the delimiters between the words in the sequence $\cal{C}$ are
indicated with the symbol $2$. In the course of the evolution of the
sequence, a number of delimiters may aggregate at successive sites,
seeming to bracket strings of zero length.  These ``null strings'' are not
associated with any nodes and ignored in the construction of the network.
The connectivity rule will be the same as that of the BE model, given in
Eq.(\ref{connmatrix}) above.

\subsection{Mutations}

At each time step $t$, for each $k$th member $x_k$ of the complete sequence $\cal{C}$, we decide with a fixed  probability $\mu$ whether it will suffer a mutation,  independently of the other members. 

\vskip 1cm
\underline{Mutation rule M1}
\begin{enumerate}
\item If $x_k \in \{0,1\}$, then we change $x_k$ to $\overline{x}_k \equiv 1-x_k$.
\item If $x_k=2$, then we {\em exchange} it with the symbol to either its right or left, with equal probability. 
\end{enumerate}

Clearly item (1) corresponds to a substitution in the code, while  item (2) corresponds to a shift in  the reading frame starting at the delimiter site. We have checked that on the average, it makes no difference whether the sites $i$ to be mutated are picked at random or sequentially. 

Extensive simulations which we reported in~\cite{Balcan-Erzan} show that the BE model is robust with respect to mutations M1. In fact, applying the set M1 a large number of times may be used to generate independent random sequences $\cal{C}$ from a given initial sequence.  Note that M1 strictly conserves the length $L$.

\vskip 1cm
\underline{Mutation rule M2}

A more complicated algorithm  M2, consists of a set of three different operations, insertion, deletion, or replacement. Once it has been  decided to mutate a site $k$, one of these operations is  picked with equal ($1/3$) probability, independently of the value of $x_k$. 
\begin{enumerate}
\item Insertion.  A character, picked with equal probability from the set $\{0,1,2\}$, is inserted to the right of the $k$th site, coming to occupy the $k+1$st site.
\item Deletion.  The $k$th element is deleted. 
\item Replacement. The $k$th element is randomly replaced by either $0$ or $1$ with equal probability.
\end{enumerate}

Note that M2 preserves $L$ on the average, since the operations of insertion and deletion are picked with equal probability, and replacement does not change the number of sites.  On the other hand, 
the insertion of the delimiter (i.e., the symbol 2) will cut a word into two successive words; the deletion of a delimiter may merge two successive words into one.  Moreover, the replacement of a delimiter by either a 0 or a 1 will again make one word out of two neighboring strings. Thus, the implementation of M2 can split or merge pairs of nodes of the underlying graph, changing the total number of nodes.  Nevertheless, the number of nodes is also conserved on the average in the long time behavior, leading to a steady state with a conserved average string length $\langle l \rangle$. 

\subsection{Duplication}

Finally,  we postulate that at each time step, a randomly chosen string $G_i$ is duplicated. The duplicated string may then be inserted in tandem with the original string, in the original or in reversed order, or inserted at a random point somewhere in the sequence $\cal{C}$.~\cite{Alberts1}  With equal probability, one of the following options is chosen:
\begin{enumerate}
\item Tandem insertion.  The copied string is inserted right after the original, in the same order.  With probability 1/2, a delimiter is inserted between the two copies, and with probability 1/2, they are run on without the delimiter (which only appears at the end of the inserted copy). Thus, with equal probability, either a new node is created, or the the word associated with the $i$th node is doubled in length.
\item Reverse tandem insertion.  Same as above, except that the polarity of the string is reversed before it is inserted.
\item Random insertion. The copied string is inserted at a random point in the sequence $\cal{C}$ without any  delimiters on either side of it. This does not give rise to a new node, but only modifies the word associated with an existing node.
\end{enumerate}

After transients, duplication leads to a steady increase in the number of strings for both sets of mutation rules. The long time, large $l$ behavior of the string length distribution can be obtained analytically for various cases, and is given in the Appendix.

\subsection{Initial conditions}

The  probability distribution according to which the initial sequence $\cal{C}$ will be formed may also be picked in different ways.
One choice, which we specified in the previous section, Eq.(\ref{probdist1}), results in an exponential distribution of string lengths, 
\begin{equation}
P_e(l) \simeq \xi \exp (-l/\xi) \;\;\;, \label{expdist}
\end{equation}
where $\xi\simeq  1/p$ for small $p$, and corresponds to a random genome. 

A second choice is motivated by considering the strings with which to associate the nodes, to stand in for the so called  promoter sequences in the gene regulatory network. In this case, inspection of the data banks publicly available on the internet~\cite{consensusseq} shows that the lengths of the consensus sequences~\cite{Lee,Harbison} are distributed in a much narrower fashion, around a central peak. This distribution we have mimicked with a Gaussian,
\begin{equation}
P_g(l) =  {1 \over \sqrt{2 \pi \sigma^2}} e^{-(l-l_0)^2/2 \sigma^2}\;\;\;. \label{Gaussian}
\end{equation}
The procedure by which the sequence is generated in this case is that the lengths $l_i$ of the successive strings are picked from the distribution in Eq.(\ref{Gaussian}), the delimiters are inserted at the sites $k_i = \sum_{j\le i}(l_i+1)$ and finally the rest of the sites filled with the symbols 0 and 1 with equal probability.

We have used the above sets of rules to construct two models, which combine them in different ways. Model I comprises the mutation set M1, with or without duplications, and with either sets of initial conditions.
Model II comprises the mutation set M2, again with or without duplications, and with either set of initial conditions.

\section{Simulation results}

In this section we summarize our simulations results for the out-degree distributions and 
the clustering coefficients for the out-neighbors. All simulations were done on sequences of initial total 
length $L_0=1.5 \times 10^{4}$.  The value of $p=0.05$ was chosen for those simulations with 
exponential initial length distributions (and therefore on the average 750 random initial sequences). 
The parameters of the Gaussian distribution are given by  $l_0= 15$ and $\sigma = 2$,  while the number of 
strings (or the number delimiters) for the Gaussian initial length distributions was $N_0=700.$  The value 
of the mutation probability was chosen $\mu=0.05$.  Averages were performed over 500 realizations in each case. 

\subsection{The out-degree distribution}

The results from the two  models should coincide for t=0, for the two different initial conditions. In Fig.1 the plot  (P) for the exponential length distribution of course reproduces the results from the original BE model, and one can clearly discern the discrete and continuous regimes.  The plot (U) for the Gaussian length distribution  shows less structure, since the paucity of very small strings shortens the discrete part of the plot, and gives it the overall appearance of just one power law distribution with $\gamma \simeq 2$, with some statistical scatter toward large degrees.

\vspace*{1cm}
\begin{figure}[h!]
\begin{center}
\end{center}
\includegraphics[width=8cm]{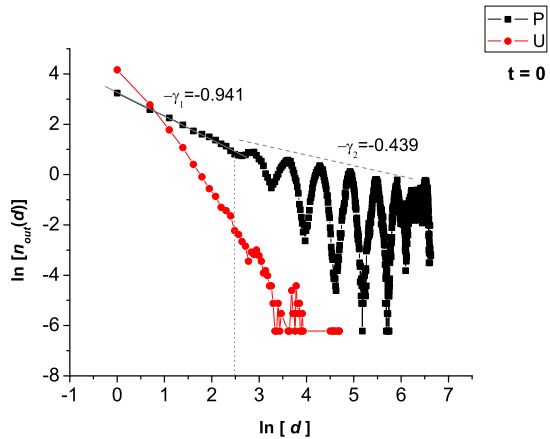}
\caption[]{(Color online)The out-degree distributions for exponential (P) and Gaussian 
(U) initial length distributions, at time t=0, shown on a double 
logarithmic plot.  For the random sequence (P), the exponent for the 
initial putative scaling region 
and the envelope for the oscillatory regime are indicated on the plot.  
The crossover between the two regimes have been indicated by a vertical 
dotted line. The Gaussian length distribution (U) gives rise to an 
overall slope of $\sim -2$.} \label{outscalingt0} \end{figure}

We should note that the length $L$ of the sequence $\cal{C}$, as well as the number of nodes, $N$, and the average word length, $\langle l \rangle$, changes  with time under duplication.  Since there is a subtle interplay between the different kinds of mutation and duplication, this change is not uniform. For Model I we  find $\langle l\rangle$, $N$ and $L$ stay constant under M1, and they all grow linearly when duplications are introduced. In Model II, $\langle l\rangle $ relaxes exponentially fast to a relatively small constant in all cases, during which time $N$ grows nonlinearly and saturates to its long time behavior.  For longer times $N$  then behaves like 
{\em i)} a constant, without duplications, and {\em ii)} grows linearly with duplications,while $L$ follows suit, with $L \sim N \langle l \rangle $.(See Appendix)

In both models, the successive application of the mutation and duplication
algorithms randomize the distribution of the delimiters within the sequence. 
Model II converges to a steady state exponential distribution of string lengths, both in the absence and presence of duplications.  Model I does so  in
the absence of duplication, where the rate is determined by the random walk step
executed by each delimiter, at each time step. (Note that double occupancy
of any site by any character is prohibited, so that the problem becomes that of
one dimensional diffusion with exclusion.) In the presence of duplication,
the population at longer string lengths grows by insertion of the copied string without the
delimiters and also with random point insertions, but the dynamics is very complicated and very persistent transients are present.  The set of
mutations M2, employed in Model II, comprising insertions, deletions
and replacements, is more effective in
randomizing the positions of the delimiters.  Below we give the results for five hundred time steps,
where subtle differences arise due to the continued presence of transients
in the simulations of sets with Gaussian initial length distributions.

\vspace*{1cm}
\begin{figure}[h!]
\begin{center}
\end{center}
\includegraphics[width=8cm]{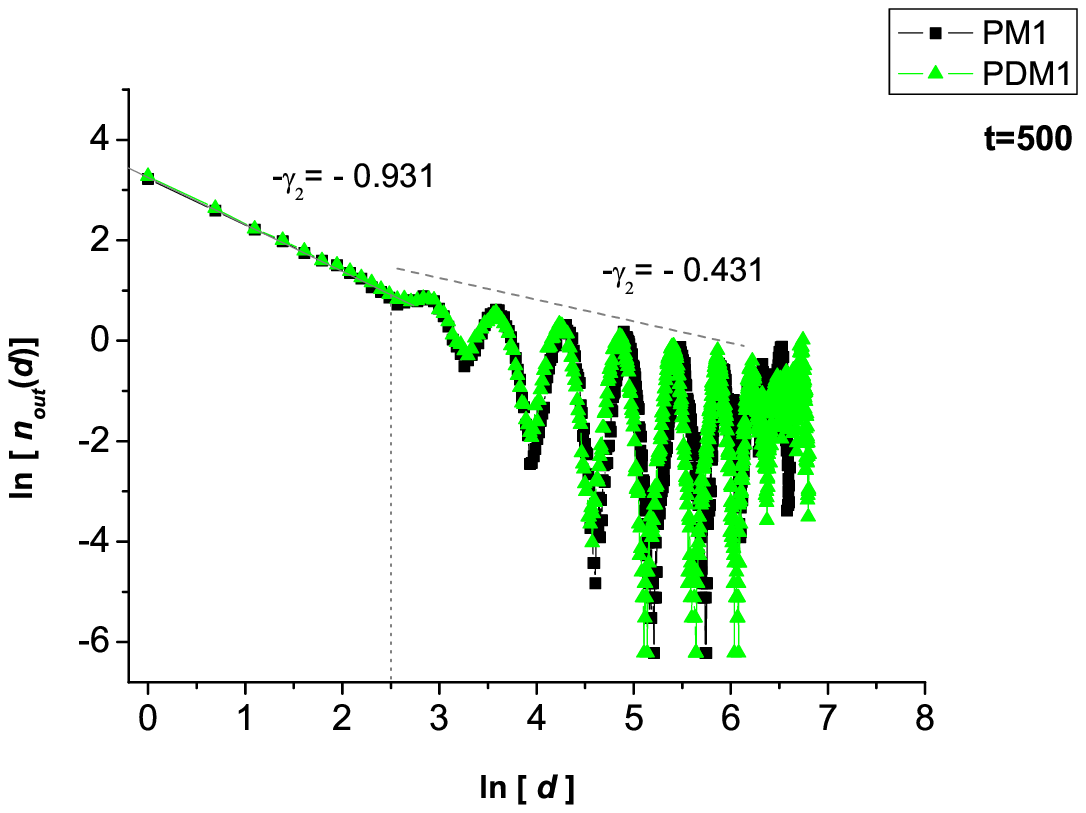}
\includegraphics[width=8cm]{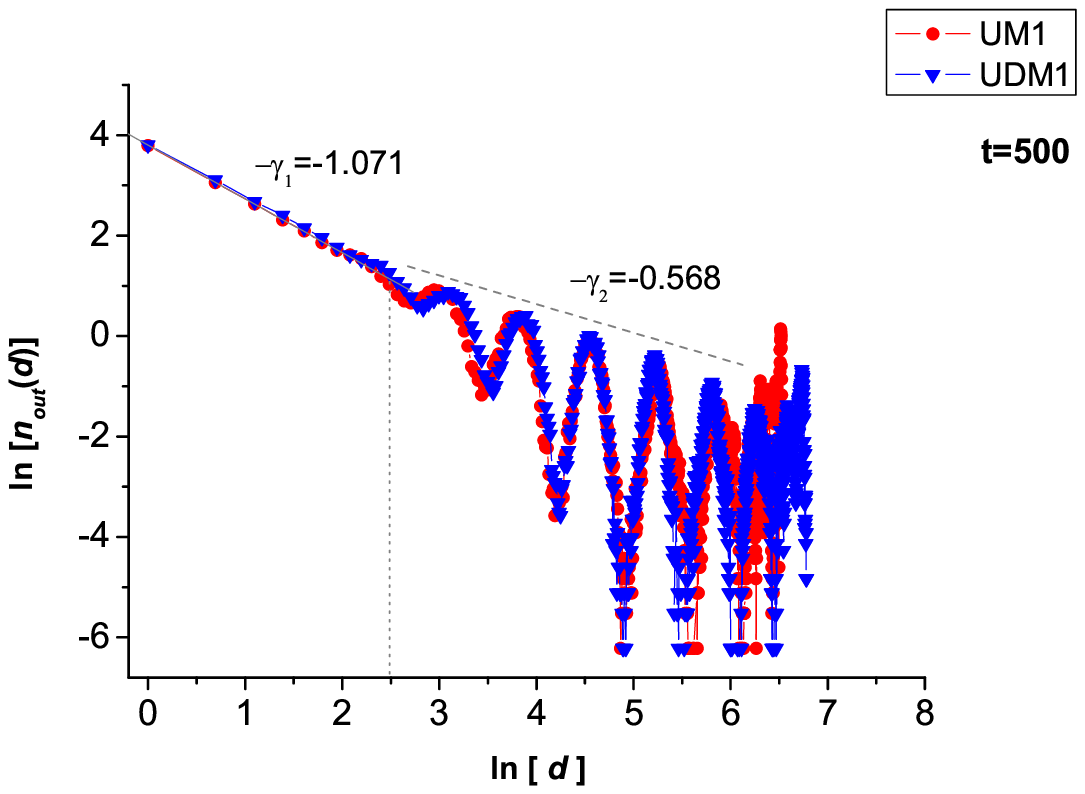}
\caption[]{(Color online) The out-degree distributions for Model I, after t=500 steps.  M1 mutation rules apply. (a) Exponential (PM1, PDM1) and (b) Gaussian (UM1,UMD1) initial length distributions, with and without duplication (D), are shown on a double logarithmic plot. See text.   }
\label{lnoutscalingIt:500}
\end{figure}

The result of 500 time steps of evolution for Model I is shown in Fig. 2.   It is remarkable that the presence or absence of duplications alters the scaling behavior very little - the plots labeled PM1 and PDM1, signifying exponential initial length distribution, M1 mutation rules, with and without duplication (D), falling almost right on top of each other. The effective $p$ value computed from Eq.(\ref{expdist}) is identical for the two graphs and the same as the initial value of $p=0.05$. The Gaussian length distributions lead to results that are almost indistinguishable, even though an inspection of the length distributions  at t=500 (not shown), reveals that they have not yet converged to exponential distributions, but show a marked peak displaced a little to the left of $l_0$, with, however a growing exponential tail for  large $l$.  The only marked difference is between the last couple of  peaks of the UM1 and UMD1 plots, seen more clearly in the  linear plots (Fig. 3) of the same quantities as in Fig. 2b.  We see that the UM1 simulation, which, without duplications, is much less deformed from the original Gaussian distribution in string lengths, and has a smaller large-$l$ tail, displays a much sharper peak at large $l$. This  corresponds to the much smaller dispersion in the out-degree distribution of the nodes with the smallest strings.

In Fig.2a, we see that the introduction of duplication does not alter the scaling exponents from
the original BE values, with $\gamma_1=0.94$ and $\gamma_2= 0.43$
falling right on top of the previously reported values for the BE model.
The exponent values for the initial Gaussian length distributions, shown in Fig. 2b,   do not show any difference with respect to the
presence or absence of duplications.  We find $\gamma_1=1.07$ and $\gamma_2= 0.57$.
\vspace*{1cm}
\begin{figure}[h!]
\begin{center}
\end{center}
\includegraphics[width=8cm]{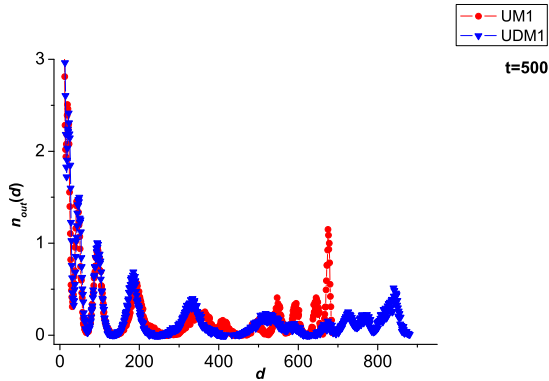}
\caption[]{The linear plot of the out-degree distribution for Model I with 
Gaussian (U) initial length distributions, and M1 mutation rules, after t=500 steps.   With (D) and without duplication.}
\label{outscalingIt:500}
\end{figure}

\vspace*{1cm}
\begin{figure}[h!]
\begin{center}
\end{center}
\includegraphics[width=8cm]{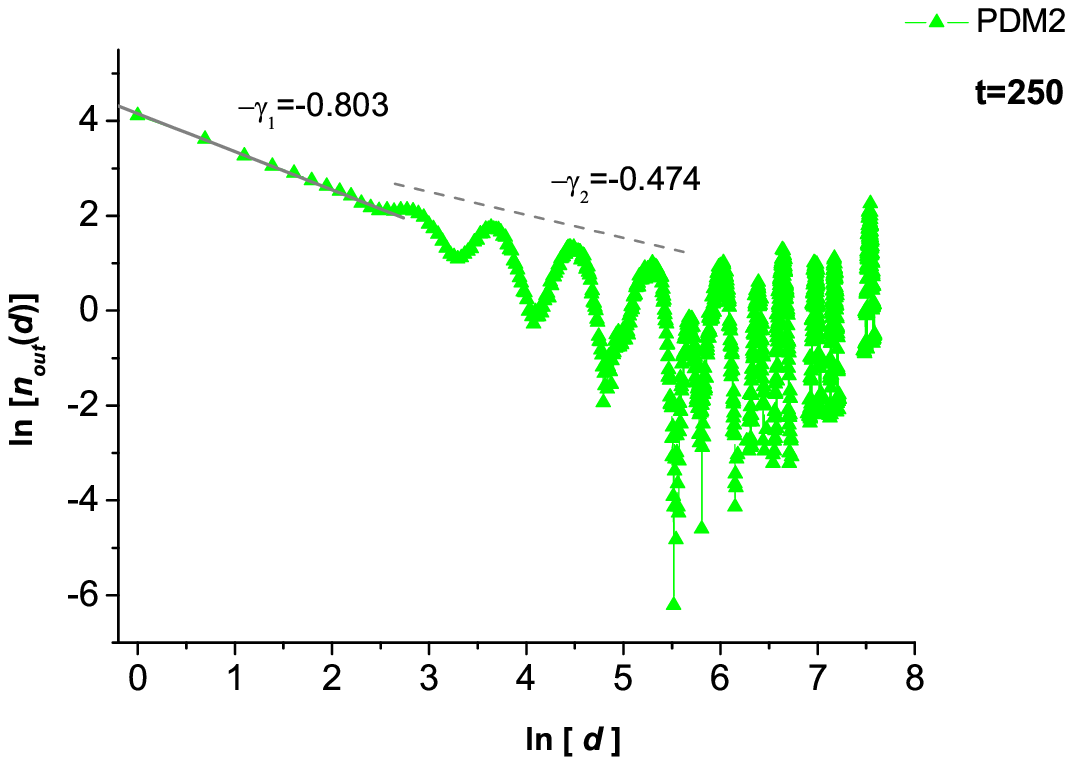}
\includegraphics[width=8cm]{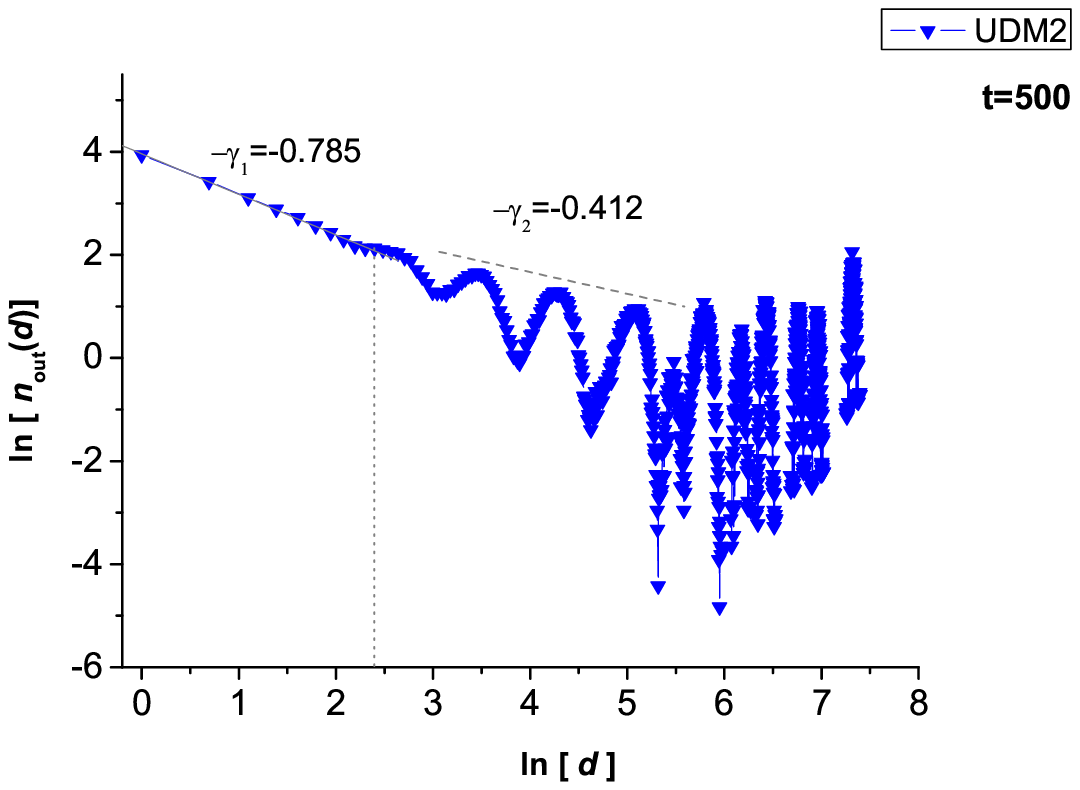}
\caption[]{(Color online) The out-degree distributions for Model II with exponential (PDM2) and Gaussian (UDM2) initial length distributions, after t=500 steps, shown on double logarithmic plots. The corresponding graphs without duplications are very similar. }
\label{lnoutscalingIIt:500}
\end{figure}

Finally, in Fig. 4, we show the scaling behavior of Model II.  The graphs
shown are very similar to those without duplications, and therefore we
omit the latter.  We observe that there is a significant change with
respect to Model I in the exponents, especially in $\gamma_1$.  The values are given in Table I.  
The first two columns are the values of $\gamma_1$ and $\gamma_2$ as
found from linear fits to the double logarithmic graphs. The third columns in the effective $p$
values found from Eq.(\ref{expdist}), from an inspection of the
exponential length distributions achieved by the random sequences.  
However, the values $\gamma^{\rm eff}$ found for the exponents upon the substitution of
$p^{\rm eff}$ in Eqs.(\ref{gamma2},\ref{gamma1}), as shown in the last two
columns, are somewhat different from the simulation results, especially for the oscillatory regimes. On the other hand, these same values,  although detectable only over a very restricted range of $\ln d$, stay rather close to the original BE results for $\gamma_2$. 

Clearly,
the assumption of statistical independence of the strings associated with
different nodes, which is employed in the derivation of these equations,
does not apply in this case.  What is remarkable, is that the mutations M2
seem to play a much stronger role in leading to correlations between nodes
of the underlying network, than do duplications.

\begin{table}
\begin{ruledtabular}
\caption{Scaling exponents for Model II after t=500 steps of mutation (M2), with or without duplication (D). P indicates a random genome, with an initial exponential distribution of word lengths, whereas U stands for an intially Gaussian distribution.}
\begin{tabular}{||c||c|c|c|c|c||}
{} & $\gamma_1$ & $\gamma_2$ & $p^{\rm eff}$ & $\gamma_1^{\rm 
eff}$ & $\gamma_2^{\rm eff}$ \\
\colrule
PM2 & {0.80} &  0.41 &  0.147  & 0.81 & 0.31 \\
UM2& 0.76 &  0.43 &  0.141  & 0.82 & 0.32 \\
PDM2& 0.80 &  0.47 &  0.145  & 0.82 & 0.32 \\
UDM2& 0.79 &  0.41 &  0.14  & 0.82 & 0.32 \\
\end{tabular}
\end{ruledtabular}
\end{table}

In Fig. 4 and 5, we see that the increase in the numbers of nodes  combines with the relatively short average string length in Model II, to remove the  finite size effects which dominate the behavior observed for the BE model and for Model I. As shown by an  explicit calculation (~\cite{mungan} and Section 2), that the high out-degree distribution for the infinite sequence, i.e., for the case $L \to \infty$, the hight of the peaks for successive out-degrees should in fact increase with $d$, and this is what we observe in Fig. 4 and Fig. 5. 

For the very large values of the degree $d$, we observe another effect
which can be called the fine-structure of the degree distribution, which
has been discussed by Bilge et al.~\cite{Bilge}, who have shown that the
exact probability for a sequence of length $l$ to be found in another
sequence of length $l^\prime$ is determined by a number called the ``shift
match number'' of the first sequence. (For any given binary sequence of
length $l$, representing a number $a$, the shift match number $s(a)$ is
given by the binary sequence $s_i(a)$, $i= 0,\ldots, l-1$, where $s_i(a)=
1$ if the subsequence $a_{i+1}, a_{i+2} \ldots a_l$ is congruent to $a_1,
a_2, \ldots a_{l-i}$, and zero otherwise.) Each of the peaks corresponding
to different string lengths $l$ split into distinct peaks according to the
shift match numbers, as the number $N$ of strings in the sequence
$\cal{C}$ becomes very large, so that the true degree distributions of at
least the very short strings are sufficiently sampled. The leading peak on
the right in Fig. 5 corresponds to $l=1$, the twin peaks on its left
correspond to $l=2$, while the next three peaks correspond to $l=3$.

\vspace*{1cm}
\begin{figure}[h!]
\begin{center}
\end{center}
\includegraphics[width=8cm]{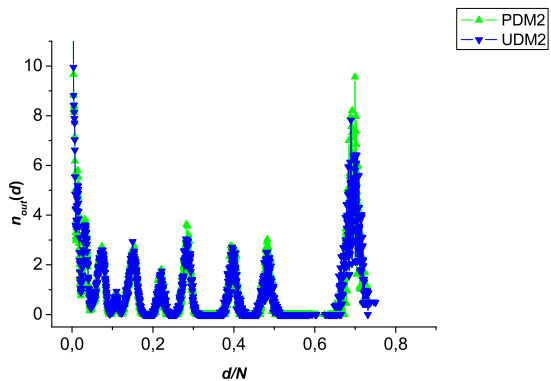}
\caption[]{(Color online) The out-degree distributions for Model II with duplications, after t=500 steps, shown on a linear plot. We have plotted $d/N$ on the horizontal axis, to bring the networks with different $N$ into coincidence. }
\label{outscalingIIt500}
\end{figure}

\subsection{The clustering coefficient $C_{\rm out}(d)$}

\begin{figure}[h!]
\begin{center}
\end{center}
\includegraphics[width=8cm]{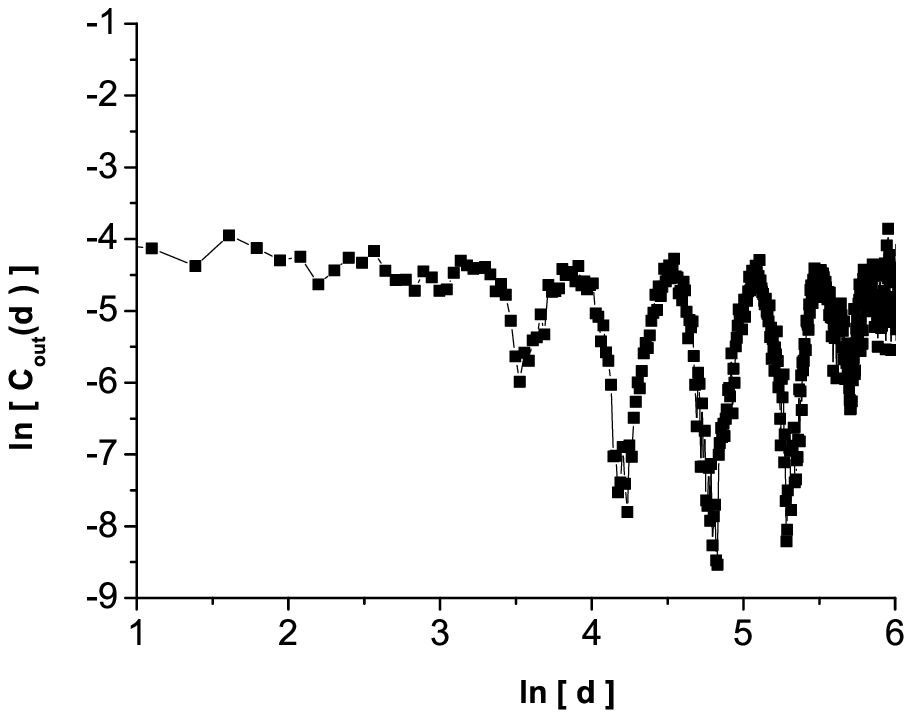}
\includegraphics[width=8cm]{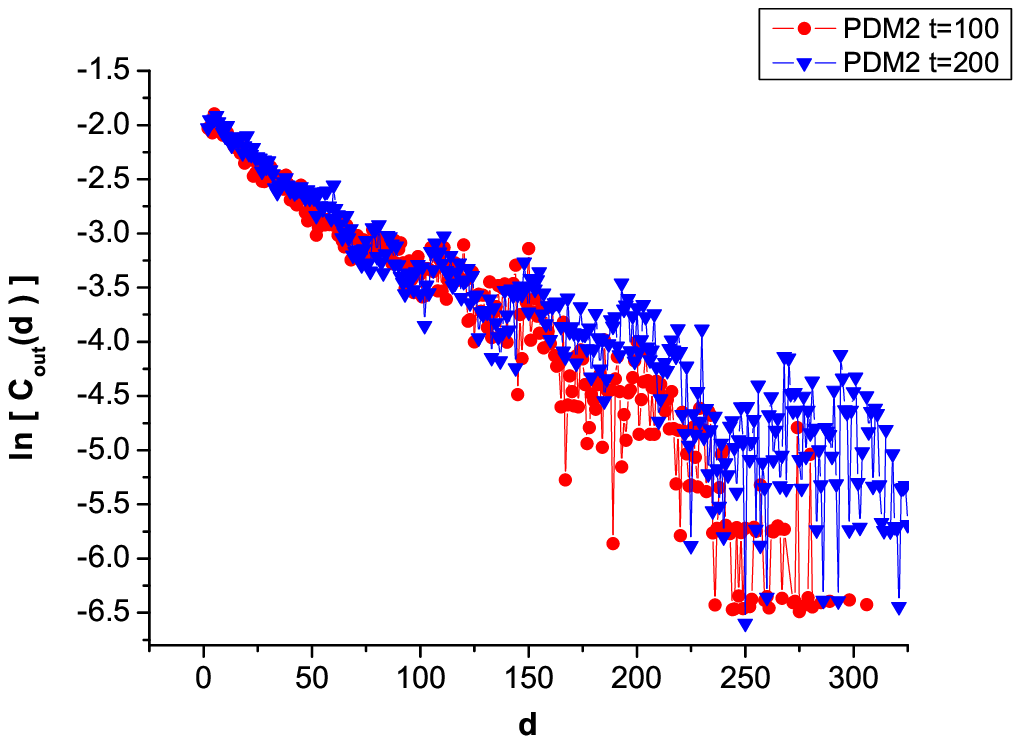}
\caption[]{(Color online) The out-clustering coefficient for the BE model (upper panel) and  Model II,lower panel, with duplications and mutations. The results have been averaged over 200 realizations. }
\label{outclusteringIIt200}
\end{figure}

In Fig.\ref{outclusteringIIt200}, we have plotted the out-clustering 
coefficient $C_{\rm out}(d)$, 
Eq.(\ref{out-cluster}), as a function of the out-degree. We see that our 
analytical calculation for the BE model is born out very well, with the 
slopes of either the initial continuous regime, or the envelope of the 
oscillatory region, being essentially negligible.  In 
Fig.\ref{outclusteringIIt200}b we report 
the results for Model II, for which the exponents for the out-degree distribution show a quantitative departure from those of the BE model. Once we turn on the mutations and duplications, the behavior is changed qualitatively, to an exponential decay with $d$, exhibiting an effect which one may call ``out-neighbor repulsion.''

\vspace*{1cm}

It is possible to gain a heuristic understanding of this exponential decay, or repulsion, by considering the fact that in Model II, at each time step one string out of $N$ is duplicated (the copy automatically inheriting all previous connections by our definition of connectivity), while strings at both ends of a bond, say of lengths $l$ and $k$ will suffer, on the average, $\mu l$ and $\mu k$ mutations.  Clearly the mutations are driving the strings apart at a faster rate than they are being duplicated.  For a hub of length $l$ to stay connected to a string of length $k$ after one time step, one needs $l/k \ll 1$, so that those mutations that do occur have a greater probability of taking place outside of the matching zone. On the other hand, for out-neighbors of length $k_1$ and $k_2$ to stay connected to each other becomes less probable, the longer they are, since they will  suffer mutations with greater probability.  This drives the connected clusters to have hubs with relatively small $l$, with out-neighbors of commensurable lengths $k \gg l$.  

In the discrete region of the out-clustering coefficient, which is still vaguely identifiable in Fig.6,  it is easier to see what is going on. By going back to Eq.(\ref{out-cluster-3}), we observe  that the lower limits of the sums must be effectively  replaced by some $\Lambda \gg l$, according to the above argument.  Moreover, in Eq.(\ref{ratio}), the numerator can be well approximated in this limit by $p(k_1,k_2) \simeq (k_2-k_1+1)z^{k_1}$, while in the denominator, we have $p(l,k_2) = 1-(1-z^l)^{k_2-l+1} \simeq 
\exp[-(1-z^l)^{k_2-l+1}]$, where we have used the fact that the expression in the parenthesis is reasonably small for $l$ small. This expression raised to $k_2-l \gg 1$ becomes much smaller than unity, so that the series can be exponentiated.  Doing this, and once more employing the trick of substituting, in the resulting expressions, the $l$ associated with  the peak values of $d$, namely $l = \ln (pd/N) / \ln (qz)$, we find that 
\begin{equation}
C(d) =f_\Lambda(q,z) e^{-\zeta d^{\gamma^{\rm eff}_1}}\;\;\;,\label{out-son}
\end{equation}
where $\zeta \equiv (p/N)^{\gamma_1^{\rm eff}} \Lambda$ and the  effective value of 
$p$ (see Table I) should be used in $\gamma^{\rm eff}_1$.  The predicted 
behavior is not exactly exponential, but with $d$ in the exponential 
having a power close to unity. Although this argument is not at all 
rigorous, if we estimate $\zeta$ from Fig.\ref{outclusteringIIt200} to be 
$O(10^{-2})$, and using 
$p^{\rm eff} = 0.15$,  $\langle N \rangle =2500$ at $t=200$ from our simulations,  we find $\Lambda=\zeta (p/N)^{-\gamma^{\rm eff}_1}\simeq 24$, which would be very much in the right ballpark, for $l \sim O(1)$.

\section{Discussion}

In this section we would first like  to give a brief overview of some selected papers where different artificial genome models have been studied.  We then summarize our results.

Geard and Wiles~\cite{GeardWiles} have used the AG proposed by Reil~\cite{Reil}, with a four letter alphabet, incorporating genes, TFs, and binding sites, or regulatory sequences. In their version, the mRNA reproduce a beginning segment (of given length) of the genes, with complementarity being interpreted as identity. The mRNA is translated to a shorter sequence (an artificial protein, or TF), in a many-to-one mapping,  once more in the same 4 letter alphabet.  The network is established by matching the proteins, which are all of the same length, to subsequences to be found within the regulatory sequences upstream of the genes and separated from them by delimiter sequences (the so called TATA boxes). They have also introduced so called sRNAs (short RNA's which bind directly onto the regulatory sequences)which enable the creating of hubs with connections to many different genes. 
Their results regarding the degree distribution of the regulatory network are not very realistic, in spite of the great pains they have taken to model the various levels of transcription and matching in a realistic way.  They find that the in-degree distribution is exponential, while the out-degree distribution is also exponential but with superposed oscillations reminiscent of those found in the BE model and the present study.  The reason they are able to observe these oscillations coming from the discrete matching sequence lengths, is because they average over many realizations of the artificial genome. 

In the paper where Watson, Geard and Wiles~\cite{Wiles} implement duplication and mutation with biologically convincing mutation operators on the AG of Reil, they unfortunately do not report on the degree distribution, but only the average connectivity and clustering coefficient of the genomic network.  Predictably, these quantities do not vary much as the network evolves, except that there seems to be a small downward trend in the average clustering coefficient.

Finally, van Noort, Snel and Huynen~\cite{vannoort} have proposed a model, where they use duplication and mutation to generate an AG from a small initial pool of 25 genes and random TF binding sequences (TFBS).  In their model, the genes and the TFBS's may be duplicated by either themselves or in pairs, after which they diverge by independent mutations. The duplicated TFBSs may be  randomly inserted upstream of other genes. The TFBSs are of fixed length, shorter than the genes. These authors have considered the coexpression network, constructed from sequence matching between  TFBSs.  The degree distributions for various relative frequencies of duplication and mutation obtained in this case are extremely similar to the BE results, with a scaling exponent of $\gamma \simeq 1.5$, except that they report only single realizations, rather than the expected distribution (averaged over many realizations), and therefore no oscillations are observed in their graphs.  

Banzhaf~\cite{Banzhaf} has considered a binary AG, the network being established by complementarity between the protein (TFs) produced by one gene, and the TFBS of another gene, both sequences being of the same length, and the AG being subjected to duplication and divergence through mutations.  Instead of the degree distribution, Banzhaf has examined the frequency of occurrence of different three-node motifs.  It is extremely interesting that the duplication-divergence procedure prunes the distribution obtained for the random AG to bring it into much closer resemblance with those obtained for the {\it Escheriche coli} and yeast genomes.

We have shown that constructing a content based version of the Wagner~\cite{Wagner,Wagner1} model by introducing duplications as well as mutations into the BE model~\cite{Balcan-Erzan,mungan} yields a complex network which is rather close in its topological properties to the original BE model. The exponents in the putative scaling region observed for small $d$ change little from their original values, staying near unity.  The behavior in the second, oscillatory regime, is enriched due to the growth of the number of nodes, resulting in the removal of finite size corrections for very small $l$, i.e., very large $d$.

In the context of our content-based model, it was appropriate to introduce an out-clustering coefficient defined as the probability that the out-neighbors of a node are connected among themselves.  We were able to give analytic derivations of the dependence of $C_{\rm out}(d)$ on the degree $d$ of the node, for the BE model.  These results were supported by our simulations.  It should be noted that all the triangles contributing to this clustering coefficient which we have defined, are in the form of ``feed-forward loops,'' motifs which are believed to play a distinctive role in gene regulatory networks~\cite{Milo}.  In the presence of duplications and mutations, numerical simulations showed that the putative power law dependence on $d$ gave way to an exponential decay.  More work is needed to explore the whole phase diagram with larger duplication rates in comparison to the mutation rates.  The numbers that were chosen here were motivated by biologically observed~\cite{Wagner,Alberts} rates of duplication and mutation.

\vskip 1cm

{\bf Acknowledgements}
 This work was done while AE was a fellow of Collegium Budapest, Institute for Advanced Study, during the 2004-2005 academic year. She would also like to   acknowledge partial support from the Hungarian Academy of Sciences and the Turkish Academy of Sciences.  YS would like to acknowledge support from the MRE fund.

\vskip 1cm
{\bf Appendix}

Here we would like to write down the equations for the  evolution of the length distribution, $p(l,t)$, under different mutation rules, in the presence or absence of duplications, and present solutions for the long time, steady state behavior.

It is useful to make a number of definitions. Let $L$ be the total length
of the sequence $\cal{C}$, consisting of the symbols $\{0,1,2\}$. We will denote
the symbol 2 as a delimiter, and those uninterrupted sequences of
$\{0,1\}$ as ``strings.'' Let $N$ be the number of times that the
delimiter occurs in $\cal{C}$.  The effective density of the delimiters is given
by $p_e=N/L$.

If the delimiters are distributed over $\cal{C}$ in a random fashion, the probability that a delimiter is followed by a string of length $l\ge 0$ is given by 
\begin{equation}
p(l)= p_e \; (1-p_e)^l \;\;,
\label{exponential}
\end{equation}
which is correctly normalized to unity if we neglect the contributions
coming from the finite size, and effectively take $L \to \infty$.  
Clearly $p(0)=p_e$. Note that if, instead of summing the discrete series,
one goes to the continuum limit and integrates over $l$, then one obtains
a slightly different normalization, with $p_{\rm cts.}(l) = \alpha \exp (-
\alpha l)$, where $\alpha = \vert \ln (1-p_e)\vert = -\ln (1-p_e)$. The
probability of finding an $l$-string ($l\ne 0$) among all the non-null 
strings is $p_{\rm nn} = p(l)/(1-p_e) = p_e
(1-p_e)^{l-1}$, or similarly, $\alpha \exp [- \alpha (l-1)]$ in the
continuum case.

The number of $l$-strings is given by $n_l$, which obeys
\begin{eqnarray}
n_l\equiv N p(l,t) &=& L p_e \,p(l,t)\;\;\; \label{nl}\\
\sum_{l=0}^L l n_l &=& L-N = \langle l \rangle N = L(1-p_e)\;\;,
\end{eqnarray}
where $p(l,t)$ is the time dependent length distribution and should not be confused with the matching probabilities $p(l,k)$ in the text.

Finally, define  $\pi(l)$ to be  the probability that a randomly picked  site  belongs to a string of length $l>0$.
Then, using (\ref{nl}) we have
\begin{eqnarray}
\pi (l)&=&  {l n_l \over L} =  p_e \, l \, p(l) \\
\sum_{l \ne 0} \pi(l) &=& 1-p_e \;\;.
\end{eqnarray}

We now would like to write down the equations for the evolution of the
length distribution, $p(l,t)$, under different mutation rules, in the
presence or absence of duplications, and discuss the long time, steady
state behavior.  We assume that for large times, a mean-field type of
approximation can be made in writing down the master equations.

\vskip 1cm
\underline{Mutation rules M1}

This set of rules calls for choosing a site  ($i$) at random, and with probability $\mu$, 

1. exchanging  $x_i$ with $1- x_i$ 
if $x_i$ has the value 0 or 1, 

2. interchanging $x_i$ with $x_{i\pm 1}$ with equal (1/2) probability, if $x_i = 2$.

The first case does not lead to any change in $p(l,t)$.  In the second case one has the following possibilities:

a)  The delimiter happens to be the left (right) boundary of a string of length $l$.  Then, exchanging it with its left (right) neighbor will result in $l \to l+1$, unless this neighbor also happens to be a delimiter, in which case  $l$ does not change. On the other hand, exchanging the position of a left (right) delimiter with its neighbor on the right (left) will give rise to $l \to l-1$.

This gives rise to the variation of $p(l,t)$, $l>0$,  under M1,
\begin{eqnarray}
{1 \over \mu} \left. {\partial p(l,t) \over \partial t}\right)_{\rm M1}  
&= {p_e \over (1-p_e)} \{- ( 2-p_e)\, p(l,t)+  \nonumber \\  
&+  (1-p_e) p(l-1,t) + \nonumber \\
& p(l+1,t) \}\; \;,
\label{M1}
\end{eqnarray}
under the assumption that a randomly chosen site along the sequence $\cal{C}$ is occupied by a delimiter neighboring a string of length $l$ on the left (or on the right) with probability  $p_e p_{\rm nn}(l,t) $. For notational convenience we have dropped the time dependence from  $p(0,t) = p_e (t) $, the density of the delimiters on the string $\cal{C}$. 

If one chooses an initial random distribution of the delimiters  along $\cal{C}$ with a probability $p_e$, one has
$p(l,0) = p(l) = p_e (1-p_e)^l$
one finds immediately upon substituting into (\ref{M1}) that the right hand side (RHS) is identically zero, since 
the curly brackets vanish. 
Thus, (\ref{exponential}) is a stable distribution under M1.

Other initial distributions may be explored, by noting that in the large
$l$ limit, Eq.(\ref{M1}) can be rewritten as
\begin{equation}
{1\over \mu}\left. {\partial p(l,t) \over \partial t}\right)_{\rm M1}=\,
{1 \over 1-p_e} \left[ p_e^2 \,{\partial p(l,t)\over \partial l} + p_e  {\partial^2 p(l,t)\over \partial l^2}\right]\;\;,
\label{M1continuous}
\end{equation}
where it is to be understood that we mean the continuous limit of $p(l)$.
This equation is in the form of a Fokker-Planck equation with drift, on
the half line $l > 0$. The distribution (\ref{exponential}) is an
approximate stable time-independent solution of (\ref{M1continuous}), to
$O(p_e^2)$. On the other hand, if one starts with an initial (Gaussian or
delta-function) distribution centered around $l_0>0$, one can see that the
peak drifts to the origin at a rate $\mu p_e^2/(1-p_e)$, and becomes 
extremely
narrow, while the distribution eventually develops a large $l$ tail.  In
fact, (\ref{M1continuous}) may be put in the form,
\begin{eqnarray}
 {1 \over \mu} \left. {\partial p(l,t) \over \partial t}\right)_{\rm 
M1}=&
-{\partial \over \partial l} \left[ - \left({\partial \over \partial l} \,\Phi(l)\right) \,p(l,t) - p_e {\partial p(l,t)\over \partial l}\right]\nonumber \\
&\times  (1-p_e)^{-1} \;\;.
\label{FP}
\end{eqnarray}
where the effective drift potential is given by $\Phi(l) = p_e^2 \,l -b \ln (l)$, and the singular logarithmic term has been inserted to mimic the infinite barrier at $l=0$.   The steady state solution is
\begin{equation}
p(l) = b\,  l\, e^{-p_e l}\;\;,
\end{equation}
where $b$ is fixed by normalization to be $b=p_e^2$.

\vskip 1cm
\underline{Mutation rules M2}

Similar to the considerations under M1, we can write down the variation of $p(l,t)$ due to the routines of insertion, deletion and replacement under the set of mutations labeled M2.

The effect of an insertion of 0,1,or 2, with equal probability, at a random point in $\cal{C}$,  is 
\begin{eqnarray}
\left.{\partial p(l,t) \over \partial t}\right)_{\rm ins.} =& - p_e\, l\, 
p(l,t) - {2\over 3} p_e \,p(l,t)+ \nonumber \\
&+ {2\over 3} p_e \, (l-1)\, p(l-1,t)+ \nonumber \\
&+{2\over 3}  \sum_{l_1= l+1}^L p_e \,p(l_1,t)\;\;\;.
\label{insA}
\end{eqnarray}
Recall that the probability that a randomly picked site belongs to an
$l$-string is $\pi(l) = p_e \,l\, p(l)$. The first term in Eq.(\ref{insA})comes from the
insertion into an $l$-string of any of the three possible choices (0,1 or
2).  The second term corresponds to the probability of inserting a 0 or 1
into an $(l-1)$-string and the third term corresponds to the probability
of inserting a delimiter into a longer string in just the two right 
positions to end up with an $l$-string.

The effect of deleting a randomly picked character is found in a similar way,
\begin{eqnarray}
\left.{\partial p(l,t) \over \partial t}\right)_{\rm del.} &= -p_e\,l\, p(l,t) - 2 p_e (1-p_e) p(l,t)+ \nonumber \\
&+  p_e\, (l+1)\, p(l-1,t)+ \nonumber \\
&+  \sum_{l_1= 1}^{l-1} p_e p(l_1,t)p(l-l_1,t)\;\;\;.
    \end{eqnarray}
    
Randomly picking a character and replacing it with either 0 or 1 has an effect on the $l$-distribution in the event that the replaced character is a delimiter.  Thus,
\begin{eqnarray}
\left.{\partial p(l,t) \over \partial t}\right)_{\rm rep.} &=& -2 p_e p(l,t) + 2 p_e^2  p(l-1,t)+\nonumber \\ 
   &+& \sum_{l_1= 1}^{l-2} p_e \,p(l_1,t)p(l-l_1-1,t)\;\;\;.
    \end{eqnarray}
We may combine these effects in two different ways, both of which leave the average number of strings and the total length of the sequence constant, after transients.  We may, after choosing a character at random, 

a) apply insertion and deletion with equal (1/2) probability, or 

b) apply insertion, deletion and replacement with equal (1/3) probability.  

The analysis for either of the cases a) or b) follows along similar lines. One may easily show that the exponential distribution in (\ref{exponential}) always solves the steady state equations obtained in the large  $l$ limit, where, after dividing out the steady state equations by $l$,  only the coefficient of $l$ survives and has to be set equal to zero, to determine the value of $p_e$.

We obtain, for the case (a)  $p_e=1/3$ and for the case (b),  $p_e= 1/6$.  These results are fully corroborated by simulations. 

One may also combine the mutation operators with duplication. 

\vskip 1cm
\underline{Duplications}

Let $\tau$ be the probability that a duplication takes place at any given
time step. In this context, one time step is that interval of time where a
randomly chosen element of the sequence $\cal{C}$ is mutated with probability
$\mu$. It is clear that $\tau = 1/L$, since one duplication takes place
only after $L$ elements have been tested for mutation.

The probability that a randomly chosen string has length $l$, is precisely
$p(l,t)$.  The variation of $p(l,t)$ due to duplication of this string and
the insertion of the copy into the sequence $\cal{C}$, in the original or
reverse order, is $2/3 $.  With probability $1/3$, on the other hand, the
copied string is inserted at a random point of $\cal{C}$ (to the right of a
randomly picked site).  Since the insertion can be made to the right of
the left delimiter, as well as after any of the elements of the
$l$-string, it occurs with probability $(l+1) p(l,t)$ into an $l$-string.  
If, however, a copied string of length $l-l_1$ is inserted into a string
of length $l_1$, in the $l_1+1$ possible positions, the number of 
$l$-strings will be increased.  Putting all
of these terms together gives,
\begin{eqnarray}
 &&\left.{\partial p(l,t) \over \partial t}\right)_{\rm dupl.} = {2 \over 3}\,\tau\, p(l,t) - \nonumber \\
 && {1\over 3} \,\tau \left[(l+1)\, p(l,t) - 
\sum_{l_1=0}^{l-1} 
(l_1+1) \,p(l_1,t) p(l-l_1,t)\right]\;.
 \label{duplication}
 \end{eqnarray}

It is easy to see that substituting the exponential ansatz  (\ref{exponential}) in (\ref{duplication})
yields, for the RHS,
\begin{equation}
{\rm RHS}= {1 \over 3 }\tau p_e \, (1-p_e)^l \{ 1- (1-p_e) l + {1\over 2} 
p_e (l-1) (l-2) \}\;\;,
\end{equation}
where the nonlinear term in $l$ comes from the sum in (\ref{duplication}).  Thus in this case there is no possibility of making the coefficients of $l$ vanish order by order. However, since $\tau \ll \mu$ for the interesting cases, to lowest order in $\tau$ the solutions without duplication are still valid, and yield an exponential distribution.

\end{document}